# Hybrid heterostructures with superconducting/antiferromagnetic interfaces

G.A. Ovsyannikov[1,2], K.Y. Constantinian[1*], Yu.V. Kislinskii[1], A.V. Shadrin[1,2], Yu. Khaydukov[3,4], T. Keller[4], A.L. Vasiliev[5]

[1]Kotel'nikov Institute of Radio Engineering and Electronics of Russian Academy of Sciences, 125009, Moscow, Russia.

[2]Chalmers University of Technology, Department of Microtechnology and Nanoscience, Gothenburg, S-41296, Sweden

[3]Skobeltsyn Institute of Nuclear Physics, Moscow State University, 119991, Moscow, Russia

[4]Max-Planck Institute for Solid State Research, 70569, Stuttgart, Germany

[5]Kurchatov National Research Center, Moscow, Russia.

*e-mail: karen@hitech.cplire.ru

## Abstract

We report on structural, DC, X-ray and neutron studies of hybrid superconducting mesa-heterostructures with a cuprate antiferromagnetic interlayer $Ca_{1-x}Sr_xCuO_2$ (CSCO). The upper electrode was bilayer Nb/Au superconductor and copper oxide superconductor $YBa_2Cu_3O_{7-\delta}$ (YBCO) was the bottom electrode. It was experimentally shown that during the epitaxial growth of the two films YBCO and CSCO a charge carrier doping takes place in the CSCO interlayer with a depth about 20 nm. The conductivity of the doped part of CSCO layer is close to the metal type, while the reference CSCO film, deposited directly on $NdGaO_3$ substrate, behaves as Mott insulator with the hopping conductivity. The interface Au/CSCO is clearly seen on bright-field image of the cross-section of heterostructure and gives the main contribution to the total resistance of mesa-heterostructure.

## 1. Introduction

Nowadays great interest attract processes of electron transport occurring at the interfaces between superconducting (S) and magnetic (M) materials, where due to the interaction of superconducting and magnetic correlations a number of non-trivial physical phenomena occur. The exchange mechanism of ferromagnetic order tends to align spins of superconducting pairs in the same direction preventing singlet superconducting pairing [1-4]. One of the important properties of the proximity effect in a superconductor/ferromagnetic (S/F) interface is a damped oscillatory behavior of the condensate wave function induced in the F- layer. This may lead, in particular, to a π-phase shift [1-3] in the superconducting current-phase relation of S/F/S Josephson junctions experimentally demonstrated in [3]. The main attention so far has been paid to analysis of structures with a ferromagnetic interlayer [1-3], while S/AF hybrid structures with antiferromagnetic (AF) have not been studied comprehensively. The theory of S/AF/S structures was discussed in publications [4-6] where antiferromagnetic interlayer was treated as a structure composed from atomically thin magnetic F-layers with antiparallel magnetization. It was theoretically shown that the characteristic feature of the Josephson effect in S/AF/S structures is its dependence on whether the number of F-layers in AF interlayer is even or odd [5,6]. Experimentally superconducting current through an AF interlayer was observed [7,8].

However, significant part of experimental studies of S/M interface were carried out on polycrystalline M-films [3,7,8] with a reduced mutual influence of crystal structures of contacting materials. Substantially smaller coherence length of oxide superconductors than in metal superconductors considerably complicates fabrication processes of oxide superconducting structures with a magnetic interlayer. However, the anomalous proximity effect was reported in lanthanum structures [9], superconducting current through an oxide AF interlayer was observed [10, 11] for hybrid mesa heterostructures demonstrating the Josephson effect. Earlier the superconducting current was observed in cuprate superconducting junctions with an artificial barrier with a separation much larger than the coherence length [12, 13].. However, these results were interpreted in terms of occurrence of current shorts [13]. A percolation mechanism of the superconducting current passing through an anomalously thick interlayer was proposed in [14]. Recent experimental results for structures fabricated by modified advanced techniques of cuprate film growth [6, 8, 15] can not be explained by a trivial presence of pinholes.

In order to observe long range proximity effect in a superconducting structure with M-interlayer a relatively transparent S/AF interface is needed. However, in spite of promising progress in fabrication of heterostructures with the magnetic interlayer [9, 10, 15], there is still a lack of experimental data on Josephson junctions with an AF-interlayer, in particular with cuprate materials. That's why in-depth structural investigations of the interfaces composed of cuprate superconductors and antiferromagnetics are important. At the same time mutual influence of antiferromagnetism and superconductivity of cuprate with the d-wave symmetry at S/M interfaces in Josephson junctions is also necessary to uncover.

In this paper we present results of investigations of the interfaces of oxide S/M structures by means of transmission electron microscopy, X-ray, neutron scattering, and the DC properties for hybrid $S/M/S_d$ mesa heterostructures (MHS), where S is the thin film Nb/Au bilayer superconductor with s-wave order parameter, $S_d$ electrode is the cuprate superconductor $YBa_2Cu_3O_x$ with a dominant d-wave order parameter, and the M- interlayer is antiferromagnetic $Ca_{1-x}Sr_xCuO_2$ (CSCO) with $x = 0.15$ or 0.5.

2. **Experimental technique**

The CSCO/YBCO interface of double-layer epitaxial thin film was grown *in-situ* by pulsed laser ablation on (110) $NdGaO_3$ (NGO) substrate with the c-axis perpendicular to the substrate surface. Typically, the CSCO films with the thickness $d_M=10\div100$ nm were deposited on the top of 150 nm thick YBCO film. The changing of Sr content in the CSCO films was realized by varying the target composition. In order to fabricate MHS the CSCO/YBCO heterostructures were covered *in-situ* by 10 nm thick Au film and later *ex-situ* 200 nm thick Nb film by DC-magnetron sputtering in Ar atmosphere. The photolithography, reactive plasma etching and the Ar ion-milling techniques were utilized for patterning the shape of Nb/Au/CSCO/YBCO MHS. The $SiO_2$ protective layer was deposited by RF-magnetron sputtering and patterned afterwards in order to form the area of the MHS. An additional 200 nm thick Nb film was deposited on the top of the mesa for patterning of superconducting wiring. Thus, the square shape of $S/M/S_d$ MHS having areas from 10×10 up to 50×50 $\mu m^2$ were fabricated (Fig. 1). For comparison a similar fabrication steps were used for structuring of the MHS without M-interlayer [10, 16].

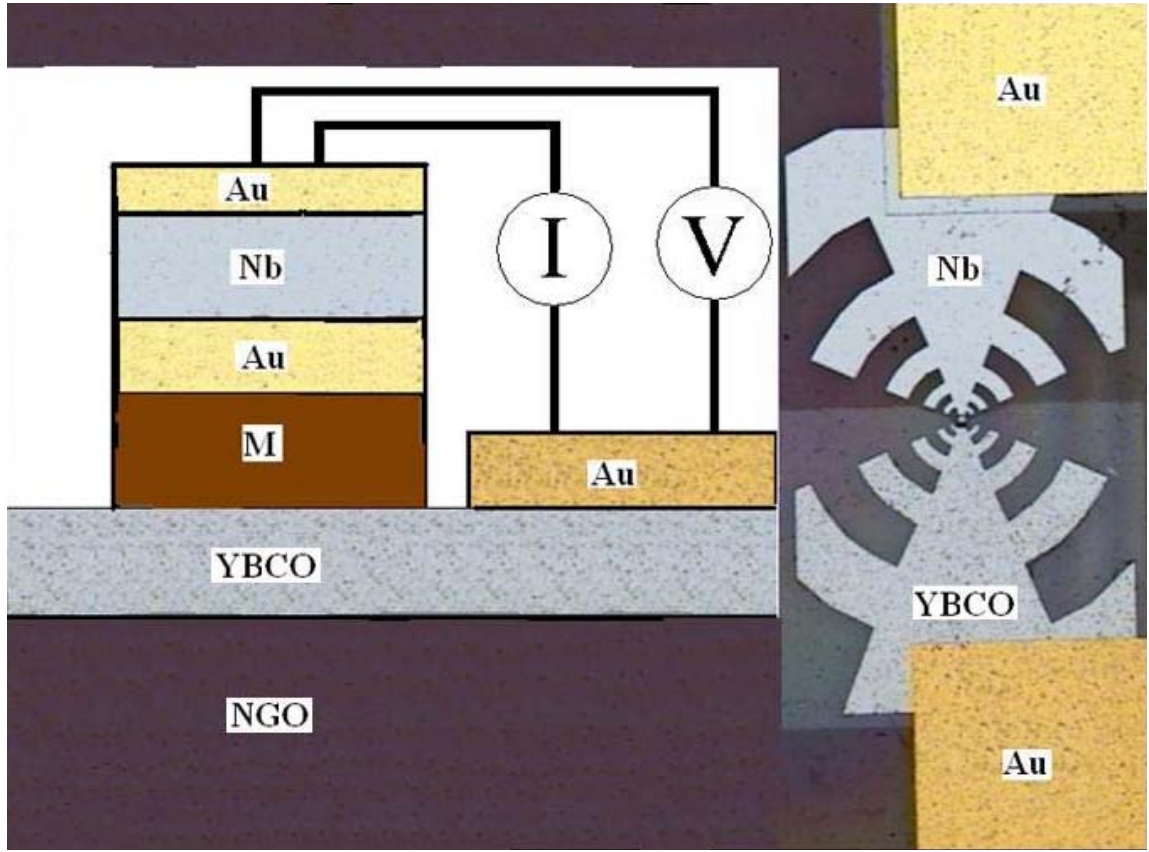

Fig.1.Chematic cross section of the MHS with magnetic interlayer and circuit for current biasing are presented on left side. The layers thickness are as following: YBCO - 200 nm, M-interlayer 5÷ 100 nm, Au- 10 ÷ 20 nm, Nb – 200 nm. Photo of MHS incorporated into log periodic antenna is presented on the right side.

The 4-point DC measurement technique was used for electrical characterization of MHS: two contacts to the YBCO bottom electrode and the two contacts to the Nb counter electrode (Fig. 1).

Structural properties of the heterostructures were characterized by X-ray (XRD) and neutron (NR) reflectometry. X-ray measurements have been conducted on the diffractometer with Cu-K$_\alpha$ ($\lambda$ = 1.54 Å) beam. X-ray 2θ-ω scan, rocking curve, the specular and off-specular reflectivity were measured. Neutron measurements have been done on monochromatic ($\lambda$ = 4.3 Å) reflectometer NREX situated on the reactor FRM II (Technische Universität München). Non-polarized beam with divergence 0.02° was used to measure specular and diffuse scattering by 200x200 mm$^2$ detector placed on 2.7 meters from the sample position.

The samples for the study of cross-sectional characteristics of heterostructure were manufactured in the electron-ion microscope Helios company using a focused ion beam energy of 30 kV at the beginning and at the end of 2 kV process. Electron microscopic studies were performed on transmit ion/scanning electron microscope (STEM) TITAN 80-300, equipped with energy dispersive X-ray microanalyses (EDRMA), EDAX, energy filter GIF, Gatan, and high-angle electron detector (Fischione) at an accelerating voltage of 300 kV.

## 3. Structure investigation

### 3.1. X-ray and neutron measurements

The X-ray scans for CSCO (x=0.15) epitaxial films deposited on the (110)NGO substrate and on the YBCO/NGO heterostructure are presented in Fig. 2. The rocking curve measurements of the Full Width at Half Maximum (FWHM) of (002) peak of the autonomous CSCO film deposited directly onto NGO substrate revealed FWHM=0.07.

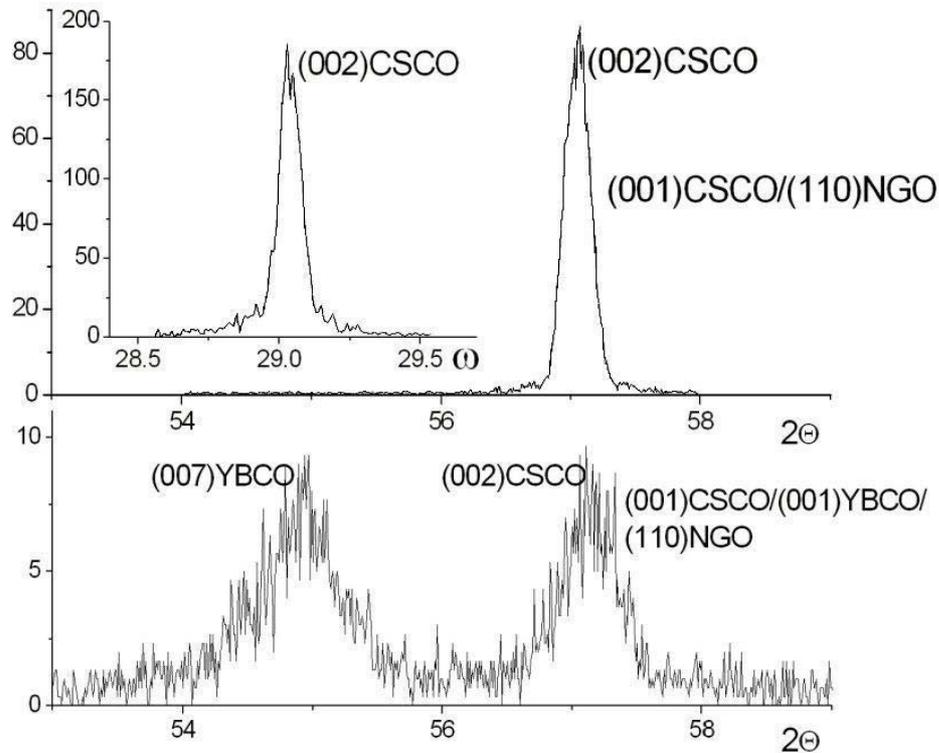

Fig.2. X-ray θ-2θ scans for an epitaxial (001)CSCO film (d=50 nm) on the (110)NGO substrate (the upper graph) and (001)CSCO/(001)YBCO/(110)NGO thin film multilayer structure (d=100 nm) (the lower graph). The rocking curve of (001)CSCO/(110)NGO bilayer is shown in the inset.

That value is smaller than FWHM=0.2° of (007) peak measured for the best YBCO film. The rocking curve measurement of the single crystal substrate (110) NGO showed FWHM=0.006° limited by resolution of the X-ray diffractometer [17]. The FWHM values of the rocking curve of CSCO film deposited on YBCO are increased by several times. All CSCO-films deposited over YBCO/NGO demonstrate a broadening of the rocking curve, manifested in a reduction of crystallographic quality and in a minor change of the lattice constants.

Since physics of proximity effects strongly depends on the quality of layers and interfaces, we studied them using X-ray and neutron reflectivity. Both methods use dependence of the scattered intensity on the space variation of the scattering length density (SLD). Depth variation of the SLD profile gives rise to the specular reflection, while presence of in-plane contrast, e.g. due to roughness causes the off-specular (diffuse) scattering. Since the neutrons scatter on nucleus and the X-rays on electron shells, the SLD profiles of neutrons and X-rays are significantly different (nominal SLDs of Au, CSCO and NGO are presented in table 1). Joint analysis of X-ray and neutron reflectivity allows to obtain comprehensive information about thicknesses of layers and the root-mean-square height, as well the in-plane correlation length of roughness on the i-th interface.

Table 1. Nominal neutron and X-ray SLDs for different materials

| Composition | density, g/cm$^3$ | Neutron SLD, $10^{-4}$ nm$^{-2}$ | X-ray SLD, $10^{-4}$ nm$^{-2}$ |
|---|---|---|---|
| Au | 19.32 | 4.51 | 123-12i |
| $Ca_{0.85}Sr_{0.15}CuO_2$ | 5 | 6.31 | 42-1.5i |
| $Ca_{0.5}Sr_{0.5}CuO_2$ | 5 | 5.47 | 40-1.4i |
| $NdGaO_3$ | 7.57 | 5.64 | 54-5.8i |

X-ray and neutron specular reflectivity curves for Au/CSCO bilayer together with the results of theoretical fit are shown in Fig. 3. The both curves are characterized by the presence of reflection plateau and Kiessig oscillations caused by the interference on different interfaces inside the structure.

In addition to the specular reflectivity, intensity of the X-ray and the neutron diffuse scattering has been analyzed to define the lateral parameters of the roughness of interfaces. Analysis has shown that the main source of the off-specular scattering is the roughness at surface with in-plane correction length of order of tens microns. Other interfaces have much smaller in-plane correlation lengths - at least two order in magnitude. As it follows from the table 1 and Fig. 3a, the X-ray reflectivity curve is mainly defined by the scattering on the gold layer.

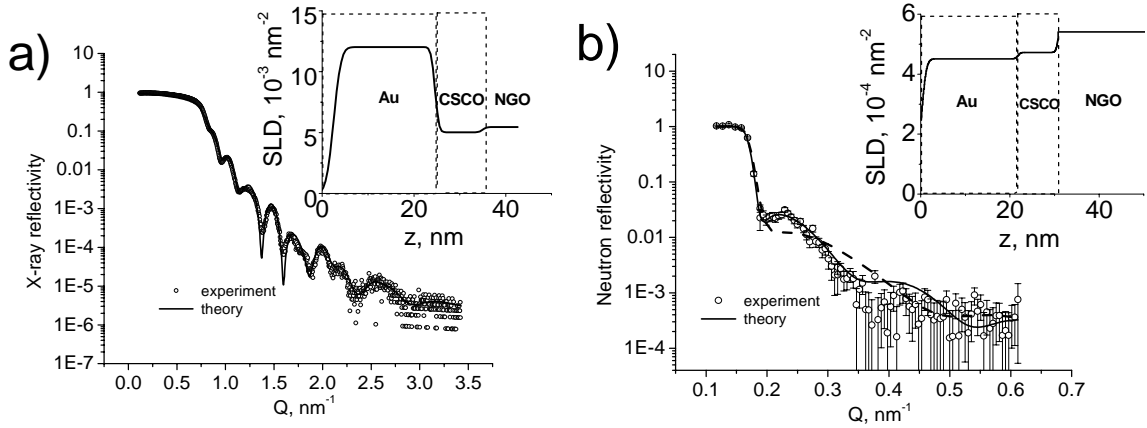

Fig.3. X-ray (a) and neutron (b) reflectivity curves (dots) with the best-fit model curves (solid lines). Corresponding SLD depth profiles are shown in the insets where dashed rectangles show the borders of the layers. The dashed line in inset corresponds to the model reflectivity curve with the nominal SLD for CSCO layer with x = 0.5.

Fit of this curve to the model allows us to define thickness of Au layer ($d_1$ = 22m) and r.m.s heights of roughness on the surface ($\sigma_1$ = 1.3nm) and on the interface Au/CSCO ($\sigma_1$ = 0.7nm). The fitting curve also shows that SLD of gold layer within 2% accuracy does not differ from bulk value (see table 1), which indicates high enough quality of the deposited gold layer. To fit the neutron specular reflectivity, thickness $d_2$ and SLD of CSCO layer the $\eta_2$ and r.m.s height $\sigma_3$ of CSCO/NGO interface were used. The SLD depth-profile, which corresponds to the best agreement of experiment and model, is depicted in the inset to Fig. 3b. It shows that CSCO/NGO interface is characterized by the r.m.s. height of roughness $\sigma_3$= 0.6 ± 0.2 nm. Thickness of the CSCO layer, according to the fit is $d_2$ = 9 ± 1 nm, which, within the error bars, equals to the nominal value. The SLD data of CSCO layer gives $\eta_2$= 4.7·10$^{-4}$ nm$^{-2}$ which is 15% less than the nominal value. In order to show the sensitivity of the fit to this parameter the model reflectivity curve for nominal value $\eta_2$= 5.5·10$^{-4}$ nm$^{-2}$ is depicted in Fig. 3b by the dashed line. This 15% difference can be explained by some decrease of the density in thin CSCO layer comparing to bulk value and/or by a change of the stoichiometry.

Thus, the analysis of neutron and X-ray specular and off-specular scattering allows us to make a conclusion about quality of the layers and interfaces in Au/CSCO/NGO system. The top layer is represented by 22 nm thick gold film with the density close to the nominal one. The thickness of CSCO layer, 10nm, is also close to the nominal one. However the neutron SLD data of the layer is 15% less than the nominal SLD for CSCO with *x* = 0.5. This discrepancy can be explained either by a decreased density of thin film or by a different stoichiometry. Analysis of the diffuse scattering proves high quality of Au/CSCO and CSCO/NGO interfaces: the root-mean-square height of the roughness of these interfaces does not exceed 1 nm. The main source of the diffuse scattering is the surface of the sample with $\sigma_l=1.3$nm and correlation length of the roughness on surface $\xi_l \cong 50$ µm.

3.2. Cross section of hybrid heterostructure.

Bright-field TEM image of the cross-section of the heterostructure Au/CSCO/YBCO/NGO without top Nb film is shown in Figure 4. There are clearly visible the interfaces YBCO/NGO and Au/CSCO.

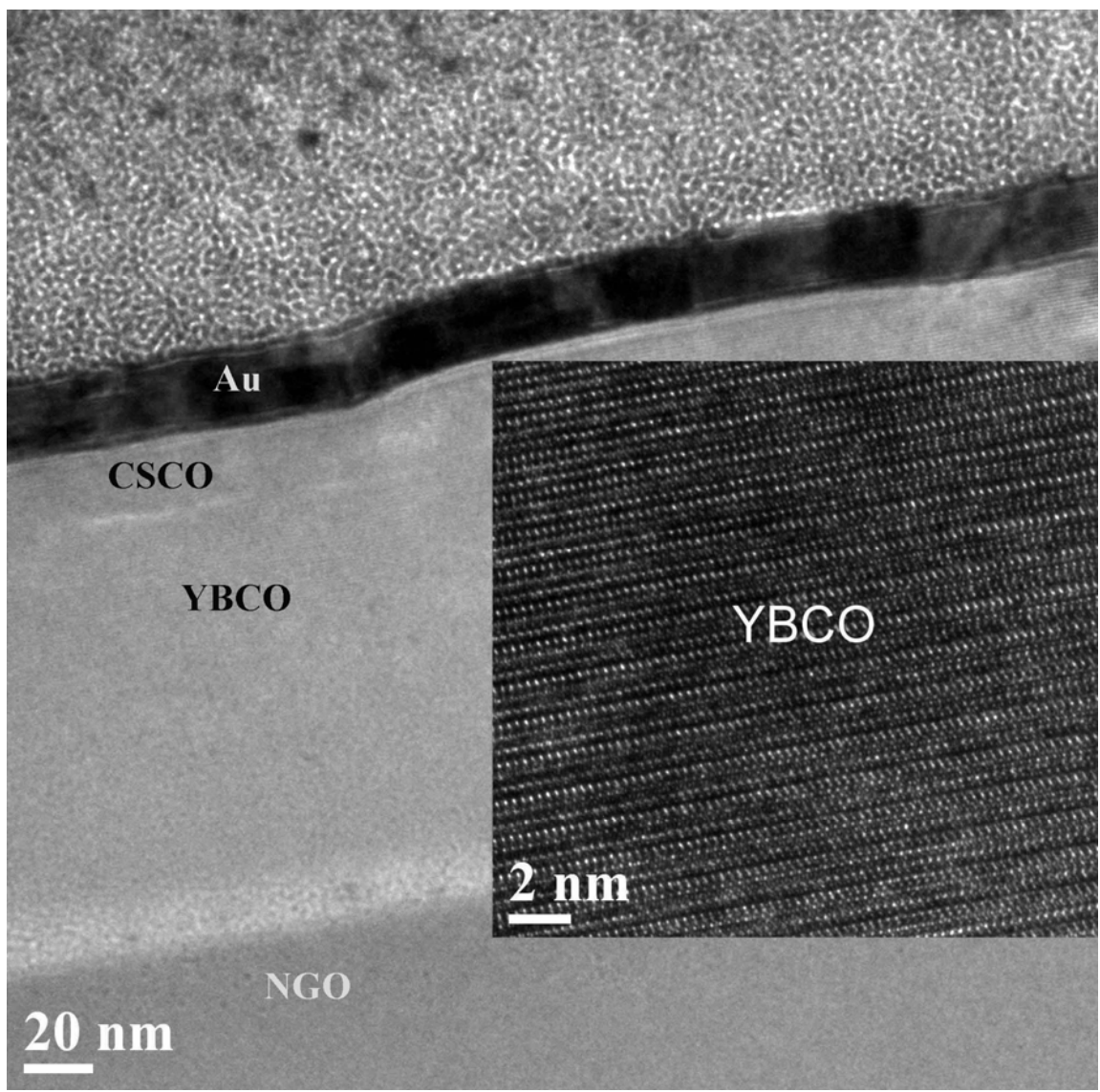

Fig.4. Bright-field image of the cross-section of the heterostructure Au/CSCO/YBCO/NGO, resulting in a transmission electron microscope. The inset on the right shows an enlarged section of YBCO films.

The inset shows an enlarged part of the YBCO film. The results of microanalysis (EDRMA) is shown in Figure 5 indicate the presence of Ca and Sr in the range of position 175 -195 nm for CSCO-interlayer. The thickness of CSCO interlayer corresponds to the estimated one from the number of laser pulses of film deposition.

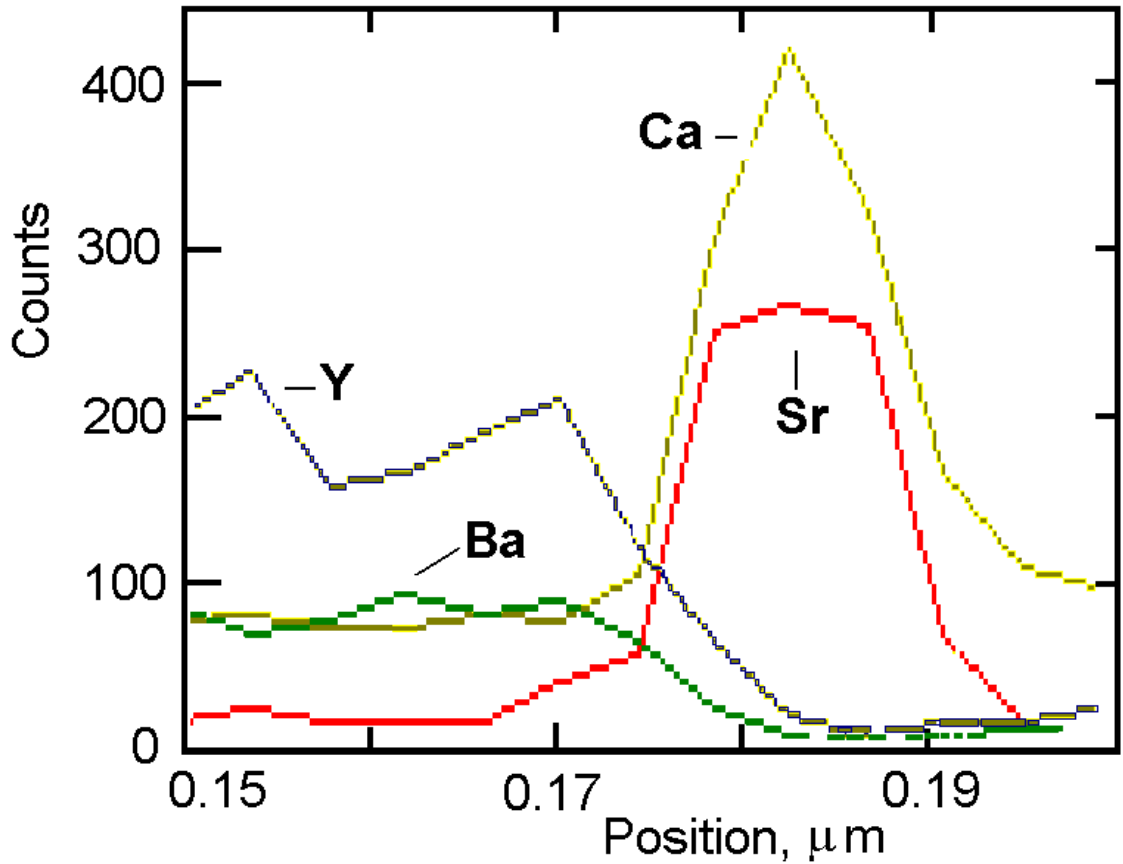

Fig.5. The results of energy dispersive X-ray microanalysis of cross-sectional part near the CSCO/YBCO interface.

## 4. DC parameters

4.1 Interface resistance

Fig. 6 shows the temperature dependence of resistivity ($\rho$) of reference CSCO films ($x = 0.15$ and $x = 0.5$) deposited on NGO substrate. In [17] it was shown that function $\rho(T)$ of reference CSCO film corresponds to a 3D - hoping conductivity [18]:

$$\ln[\rho(T)] = \ln(\rho_0)+(T_0/T)^{1/4}, \tag{1}$$

where $T_0 = 24/(\pi k_B N_F a^3)$ is experimental constant $N_F$ is the density of states at Fermi level, $a$ is the localization radius of charge carriers, $k_B$ - Boltzmann constant. For the CSCO film (x=0.5) presented at Fig. 6 we obtain $T_0 = 3 \cdot 10^6$ K and resistivity at low temperature ($T$=4.2K) $\rho > 10^4$ Ω·cm. It should be noted that in all reference CSCO films no metallic conductivity were observed.

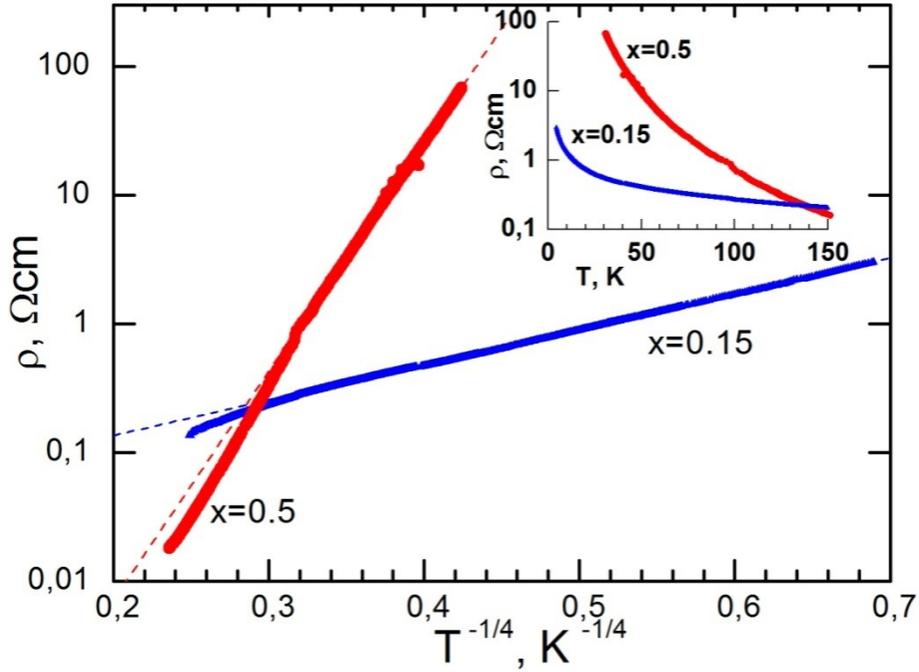

Fig. 6 Resistivity for films CSCO with $x = 0.5$ and $x = 0.15$ plotted as function $T^{-1/4}$. Dashes show 3D hopping conductivity. The inset shows the same dependences on linear scale for $T$.

Resistance, $R$, of MHS consists of sum of the resistances $R_Y$, $R_{M/Y}$, $R_M$, $R_b$, $R_{Nb}$ of the YBCO electrode, M/YBCO interface, M-layer, the interface resistance between the M and Au, Nb/Au electrode, respectively. Figure 7 shows the temperature dependence of the resistance $R(T)$ for MHS with $d_M = 20$ nm, $x = 0.5$, $L = 10$ μm. The resistance of the metal electrode Nb/Au at room temperature is small for $\rho_{Nb/Au}=10^{-5}$ Ω·cm and thickness $d_{Nb/Au} = 120$ nm. At temperatures below the critical temperature of YBCO electrode, $T<T_C$, the contribution of the metal film resistance $R_{Nb/Au}$ is small and vanished below the critical temperature of the top electrode Nb-Au ($T_C' = 8 - 9$ K). At temperatures $T>T_C$ dependence $R(T)$ of MHS is similar to $R_{YBCO}(T)$ of YBCO film measured separately. In the range $T_C'<T<T_C$ the MHS resistance $R$ is close to constant value $R_{M/Y} + R_M + R_b$.

Taking into account the epitaxial growth of two cuprates CSCO/YBCO and similar parameters of their crystal lattices, we believe that the resistance $R_{M/Y}$ of interface is small compared to $R_b$ [19]. Accordingly, on the bright-field image in Fig. 4 there are clearly visible the color contrast borders of CSCO/Au, while the boundary of YBCO/CSCO is not so distinguishable. The inset in Fig. 7 shows the thickness dependence of the $R_NA$ vs. $d_M$ for MHS, in which the Josephson effect is present. The resistance in the normal state $R_N$ was measured at the voltage $V \sim 1.5$ mV at $T = 4.2$ K, $A = L^2$ is square of MHS.

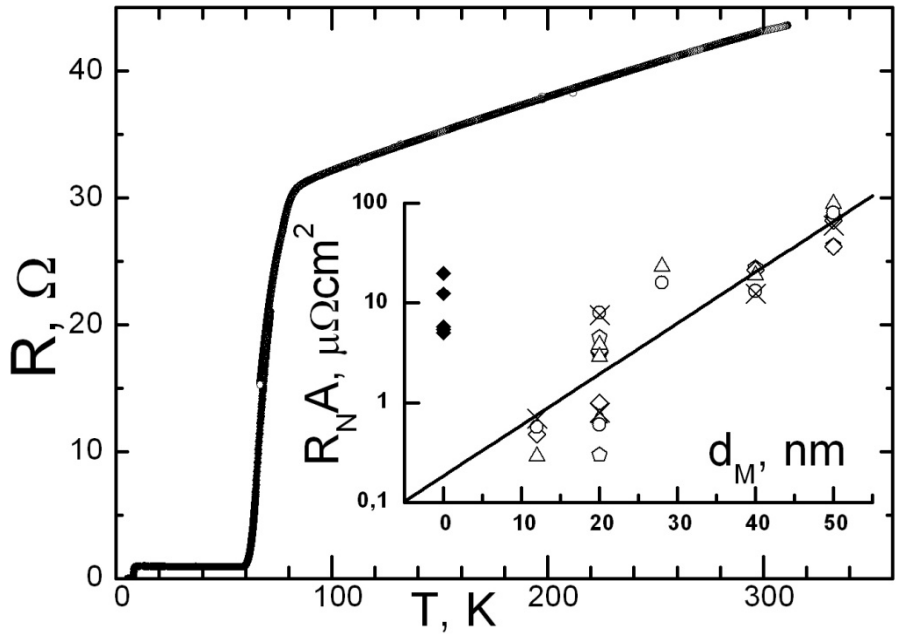

Fig.7. Temperature dependence of the mesa heterostructure with $d_M$=20 nm and $L$=10 μm. The inset shows the dependence of the characteristic resistance of $R_NA$ vs. interlayer thickness $d_M$ of CSCO with $x$=0.5 at $T$ = 4.2 K.

It is clear from the data presented in Fig. 6, that the resistivity $\rho_M$ of reference CSCO film ($x = 0.5$) increases with cooling. Since at $T = 4.2$ K $\rho_M > 10^4$ Ω·cm the expected contribution of CSCO interlayer with thickness $d_M \approx 10$ nm to MHS resistance $R_NA = \rho_M d_M > 10^4$ μΩ·cm². However, for the MHS with CSCO layer $d_M \leq 20$ nm so large $R_NA$ were not observed. Furthermore, compared with the reference CSCO film the resistance of MHS weakly depends on temperature in the range $T_C' < T < T_C$. Consequently, the main contribution to the resistance at low temperatures of the MHS with thin CSCO layer $d_M \leq 20$ nm gives the CSCO/Au interface [19]. As can be seen from the curve in the inset Fig. 7 the $R_NA$ increases exponentially with $d_M$: $R_NA = A_R \exp(d_M/a_R)$. Fitting parameters were calculated by least-squares technique: $a_R = 8.5$ nm, $A_R = 0.184$ $\mu\Omega\cdot cm^2$. These data show that for the thickness of the layer of $d_M < 40$ nm the $R_NA$ is smaller than that for structures without M-layer ($d_M = 0$). If the main contribution to MHS resistance comes from hopping conductivity in the CSCO interlayer, the $R_NA$ would linearly increase with $d_M$, which was not observed.

4.2 Interface capacitance

The additional information about the electrical properties of the boundary layer and YBCO/CSCO interface can be extracted from capacitance $C$ of MHS.

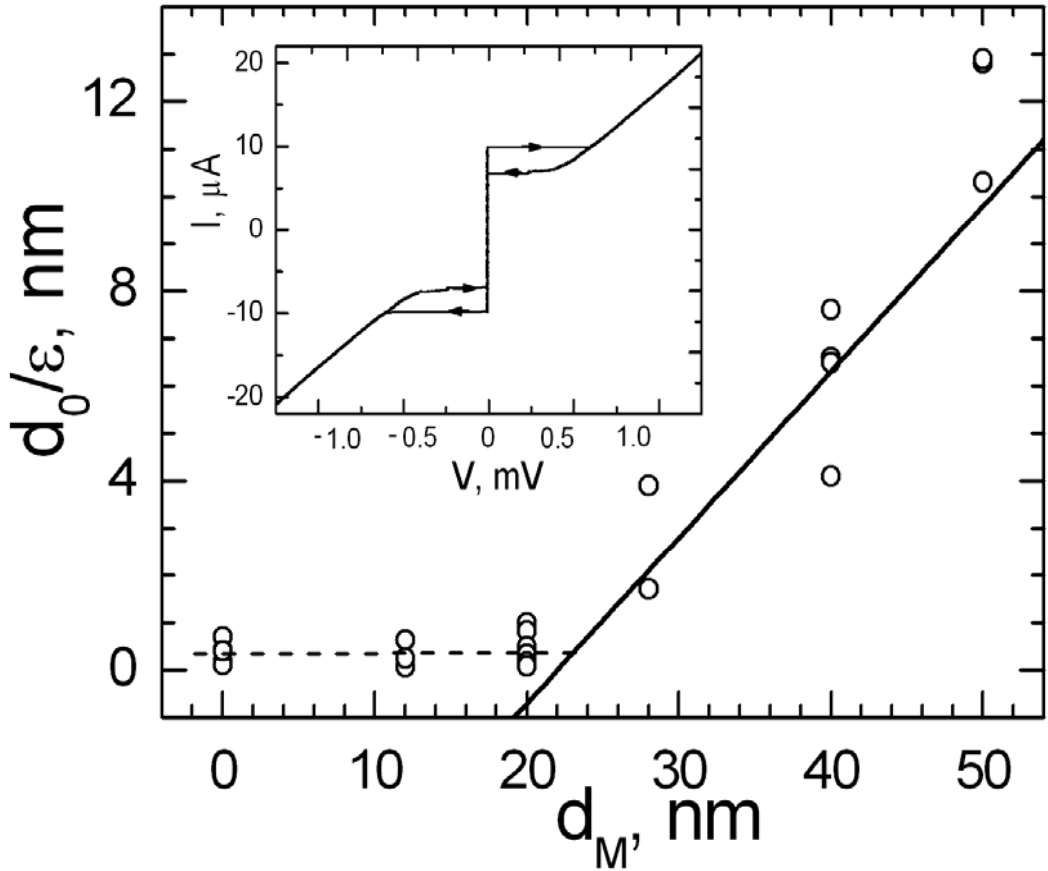

Fig.8. The thickness $d_M$ dependence of the thickness of effective barrier $d_0/\varepsilon$ in MHS for CSCO for $x = 0.5$. The dash lines show the approximation dependences. The inset shows the current-voltage characteristics of MHS with interlayer thickness $d_M = 20$ nm. The arrows indicate the critical current and the return current.

The *I-V* curve of MHS at *T* = 4.2 K has hysteresis (see inset in Fig. 8). The capacitance *C* of MHS can be found from the McCumber parameter $\beta_C = 4\pi e I_C R_N^2 C/h$ using the ratio of return current to the critical current on the *I-V* curve of Josephson junction [20]. For sandwich geometry the capacitance $C = \varepsilon_0 \varepsilon A/d_0$, where $\varepsilon_0$ - permittivity of free space, $\varepsilon$ is dielectric constant of the barrier layer with the thickness $d_0$. The dependence of $d_0/\varepsilon$ on the $d_M$ is shown in Fig. 8. For $d_M$=20 nm the ratios were $d_0/\varepsilon$=0.5±0.3 nm and had no significant difference with the value for MHS without M interlayer $d_0/\varepsilon$=0.35±0.2 nm (dash line in Fig. 8).

The presence of hysteresis on the *I-V* curve for the heterostructures without M interlayer indicates a barrier formation at the YBCO/Au interface, which determines the capacitance between the electrodes YBCO and Nb-Au. In the case of MHS with M-interlayer the barrier is formed at the CSCO/Au interface. Increasing the thickness over $d_M$ >20 nm the capacitance is decreased. Averaging by least-squares we obtain a linear dependence: $d_0/\varepsilon$=(0.35±0.05)[$d_M$–(22±4)] nm (solid line in Fig 8). The $d_0/\varepsilon$ ($d_M$) dependence is described by the model, in which a conducting layer is formed due to the influence of YBCO film on the boundary of CSCO interlayer, which one does not contribute to the capacitance *C*. At thickness above the conducting part in CSCO layer (≥ 20 nm) there is low-conducting layer with thickness $d_0$, which determines the capacitance of the MHS. Note that although the characteristic resistance $R_N A$ for MHS with and without M-interlayer differs by almost an order in magnitude, the values of $d_0/\varepsilon$ are approximately the same for MHS with $d_M$ ≤20 nm. For CSCO/YBCO interface we observe about 1 nm Sr and Ca deviation (see Fig.5) due to possible weak diffusion of cations. A similar behavior was observed at the interface of two cuprates $PrBa_2Cu_3O_7$/YBCO [21]. The changes in conductivity of the contacting materials at the interface of two oxides can be caused by electronic adjustment, as it takes place at the border of a strongly correlated Mott insulator and an insulator with a gap in spectrum of excitations [22]. Thus the contribution to resistance of MHS (if $d_M$<20 nm) does not come from CSCO/YBCO interface. It means that the main contribution to $R_N A$ gives the interface CSCO/Au, due to differences in conductivity and Fermi velocity of contacting materials, their different crystallographic parameters and the presence of defects at the borders.

For MHS with thick interlayer ($d_M$> 40 nm) a sharp change of *R(T)* is observed in temperature range $T<T_C$ as shown in inset Fig. 9. In this case, the contribution of the barrier resistance $R_b$ is small compared to the interlayer resistance $R_M$, and the resistance of the CSCO film gives the main contribution to $R_N A$ at low temperatures.

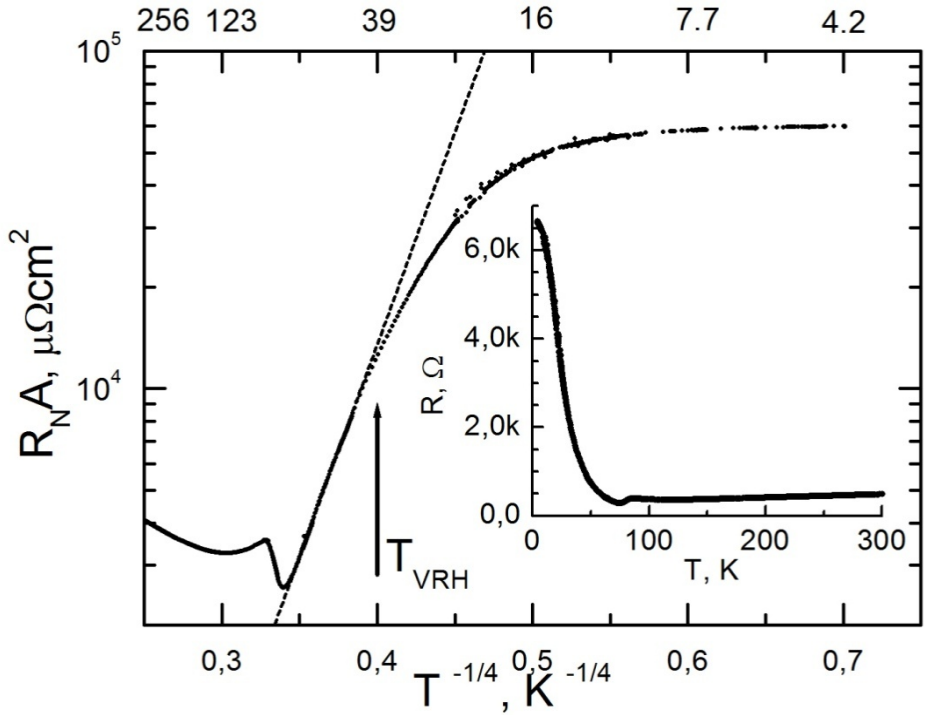

Fig.9. The characteristic resistance of MHS $R_NA$ ($x = 0.5$, $d_M = 80$ nm, $L=30$ μm) plotted as function $T^{-1/4}$. The arrow shows the crossover temperature $T_{VRH}$. Fit by (1) is shown by dashes ($T_0 = 7 \cdot 10^5$ K). The inset shows the temperature dependence (in linear scale) of the resistance of the same MHS.

In the temperature range $T = 70 \div 43$ K the $R(T)$ dependence of MHS with thick CSCO interlayer (Fig.9) is described by variable range hopping (eq. 1). Decreasing the temperature the double hopping length the $2r \approx a(T_0/T)^{1/4}$ increases and becomes compared with the barrier thickness $d_0$ at $T=T_{VRH}$ for which the change in the conductivity mechanism takes place [23]. From Fig. 9 we see $T_{VRH} \approx 43$ K, $T_0 = 7\ 10^5$ K, taking $d_0 \approx 80-20=60$ nm we estimate the localization radius: $a \approx d_0(T_{VRH}/T_0)^{1/4}=5$ nm. Using the values of $T_0$ and $a$, from $T_0 = 24/(\pi k_B N_F a^3)$ it is possible to determine also the density of states $N_F = 10^{18}$ (eV)$^{-1}$(cm)$^{-3}$, which is significantly below that $N_F$ in PrBa$_2$Cu$_3$O$_7$ [21].

## 5. Conclusion

As a result of the structural, electrical, X-ray and neutron studies of hybrid mesa heterostructures based on cuprate superconductor (YBCO) with an interlayer of a cuprate antiferromagnet (CSCO) it was found that during epitaxial growth of two cuprates the CSCO/YBCO interface is formed with a high transparency. For reference CSCO film deposited directly over the substrate, the material behaves as a Mott insulator, having hopping conductivity. At the interface of YBCO/CSCO charge carrier doping of CSCO film up to a depth of about 20 nm takes place making it close to the metal conductivity. It leads to decrease of resistivity (about of two orders) of mesa heterostructure with CSCO interlayer compared with the autonomous CSCO film. The difference in electronic parameters of Au and CSCO-interlayer determine the properties of potential barrier at the interface and it has the decisive contribution into the mesa heterostructure resistance. For CSCO-interlayer thicknesses above 40 nm the resistance of mesa heterostructures at high temperatures has dependence typical for hopping conductivity. It is possible experimentally estimate the characteristic temperature of crossover to hopping conductivity and the density of states at the Fermi level for the interlayer.

We thank I.V. Borisenko, T. Claeson, V.V Demidov, A. Kalabukhov, I.M Kotelyanskii, A.M Petrzhik, A.V Zaitsev and D. Winkler for assistance in the experiments and helpful discussions. We gratefully acknowledge the partial support of this work by the Russian Academy of Sciences, Russian Ministry of Education and Science, Russian Foundation for Basic Research 11-02-01234a, 12-07-31207mol_a, Scientific School Grant NSc-2456.2012.2, and Visby program of Swedish Institute.